\documentclass[12pt]{article}
\usepackage{amsmath}
\usepackage{amsfonts}
\usepackage{graphicx}
\usepackage{xcolor}
\graphicspath{ {images/} }
\usepackage[margin=1in,top=1.5in,bottom=1.5in]{geometry}

\begin{document}

\title{Trends, Reversion, and Critical Phenomena\\ [5pt]
 in Financial Markets}

\author{\\[20pt] Christof Schmidhuber\\ [30pt]
Zurich University of Applied Sciences\\ [0pt]
School of Engineering, Technikumstrasse 9\\ [0pt]
 CH-8401 Winterthur, Switzerland\\ [0pt]
christof@schmidhuber.ch\\  [30pt]}

\maketitle

\thispagestyle{empty}

\newpage
\begin{abstract}
Financial markets across all asset classes are known to exhibit trends, which have been exploited by traders for decades. However, a closer look at the data reveals that those trends tend to revert when they become too strong. Here, we empirically measure the interplay between trends and reversion in detail, based on 30 years of daily futures prices for equity indices, interest rates, currencies and commodities.\\

We find that trends tend to revert before they become statistically significant. Our key observation is that tomorrow's expected return follows a cubic polynomial of today's trend strength. The positive linear term of this polynomial represents trend persistence, while its negative cubic term represents trend reversal. Their precise coefficients determine the critical trend strength, beyond which trends tend to revert.\\

These coefficients are small but statistically highly significant, if decades of data for many different markets are combined. We confirm this by bootstrapping and out-of-sample testing. Moreover, we find that these coefficients are universal across asset classes and have a universal scaling behavior, as the trend’s time horizon runs from a few days to several years. We also measure the rate, at which trends have become less persistent, as markets have become more efficient over the decades.\\

Our empirical results point towards a potential deep analogy between financial markets and critical phenomena. In this analogy, the trend strength plays the role of an order parameter, whose dynamics is described by a Langevin equation. The cubic polynomial is the derivative of a quartic potential, which plays the role of the energy. This supports the conjecture that financial markets can be modeled as statistical mechanical systems near criticality, whose microscopic constituents are Buy/Sell orders.\\ \\


\end{abstract}

\thispagestyle{empty}
\newpage
\setcounter{page}{3}

\section{Introduction}

It is well-known that financial markets across all asset classes exhibit trends. These trends have been exploited very successfully by the tactical trading industry over the past decades, including the former "turtle traders" \cite{turtles} and today's CTA industry.\\

A close look at the available data reveals that those trends tend to revert as soon as they become too strong. In this paper, we demonstrate this based on 30 years of daily futures returns for equity indices, interest rates, currencies and commodities. We analyze trends with 10 different time horizons, ranging from 2 days to 4 years, and empirically measure the critical strength, beyond which trends tend to revert. Here, the "strength" of a trend is defined in terms of its statistical significance, namely as the $t$-statistics of the trend. \\

In a first step, we measure the daily average return of a market as a function of the values of the 10 trend strengths on the previous day. In order to increase the statistical significance of the results, we aggregate across different markets and time scales. Our key observation is that tomorrow's average return can be quite accurately modeled by a polynomial of today's trend strength. It consists of a positive linear term that is responsible for the persistence of trends, and a negative cubic term that is responsible for the reversion of trends. Trends tend to revert beyond a critical trend strength, where the two terms balance each other. The corresponding regression coefficients are small, but statistically highly significant.  \\

In a second step, we refine this quantitative analysis. Using multiple nonlinear regression, we empirically measure how the observed cubic function varies
\begin {itemize}
\item with the time scale of the trends: we find that trends of medium strength persist at scales of several days to several years, while reversion dominates at shorter or longer time scales. We model this scale-dependence by polynomial regression as well.
\item with the asset class: we find that the available data do not allow us to fit different model parameters to different asset classes. Within the limits of statistical significance, the model parameters are thus universal, i.e., independent of the asset. 
\item over time: we find that the patterns have gradually changed over the decades. In particular, trends have become less persistent, and there is little evidence that classical trend-following can perform as well in the future as it did in the past.   
\end{itemize}
Since financial market returns are only in a rough approximation independent, normally distributed random variables, we cannot trust the standard significance analyses for regression results. Instead, we use bootstrapping and cross validation to confirm that our results are statistically highly significant out-of-sample, and robust. Throughout this paper, we try hard not to introduce a single parameter more than is absolutely necessary to capture the essence of the empirically observed patterns. We find that we may fit at most 6 parameters to our 30-year data set, and identify what we believe are the 4-6 most relevant parameters. \\

While trends have been exploited by the systematic trading industry for decades, they arrived relatively late in academia. Early observations
on market trends appear, e.g., in \cite{cutler,silber}. Early literature on the interplay of trends and reversion has focused on their cross-sectional counterparts (momentum and value) for single stocks \cite{asness}. With the advent of alternative beta strategies \cite{fung,jaeger}, trend-following has become an active academic research area \cite{miff,shen,mosk,menk,hoyle,bal}. By now, there is an extensive literature on trend-following, including backtests of its performance more than a century into the past \cite{lemp,hurst}, and efforts to optimize trend-following strategies by machine learning methods \cite{lim}. For a recent review of trend- and reversion strategies, see \cite{baz} and references therein.\\

Much of the financial literature in this field tries to improve trading strategies, be it by new trend signals, by new algorithms for mapping signals to position sizes, by identifying market environments in which a given strategy works best or worst, or by reducing trading costs or risks. However, while the results reported in our article also have implications for investors (e.g., they signal when to exit trends), our key motivation for publishing them goes much further: as discussed in section 5, the cubic polynomial, the scaling relations, and the universality that we observe all point towards a potential deep analogy between financial markets and statistical mechanical systems near second-order phase transitions. This in turn supports the idea that markets can be modeled in terms of "social networks" of traders. Our results lay the empirical basis for systematically analyzing the nature of these networks.\\

As a corollary, our observations also support a modified version of the efficient market hypothesis: they suggest that market inefficiencies do exist, but disappear before they become strongly statistically significant. In addition, our measurements quantify how markets have become more efficient with respect to trends over the decades.

\section{Data and Definitions}

\subsection{Data}

Our analysis is based on historical daily log-returns for the set of 24 futures contracts shown in table 1. This set is diversified across four asset classes (equity indices, interest rates, currencies, commodities), three regions (Americas, Europe, Asia) and three commodity sectors (energy, metals, agriculture). We use futures returns, instead of the underlying market returns, because futures returns are guaranteed to be marked-to-market daily. Moreover, they are readily available for all asset classes and net of the risk-free rate, which also makes returns in different currencies and interest rate regimes comparable with each other. \\

\begin{tabular}{ |p{3.3cm}||p{3.3cm}|p{3.3cm}|p{3.3cm}|  }	\hline
	{\it Table 1: Markets}& America  &Europe &Asia\\	\hline\hline
	Equities   & S\&P 500    &DAX 30&  Nikkei 225\\
	&  TSE 60  & FTSE 100   & Hang Seng\\ \hline
	Interest rates &US 10-year & Germany 10-year &  Japan 10-year \\
	&Canada 10-year  & UK 10-year &  Australia 3-year \\	\hline
	Currencies&   CAD/USD  &  EUR/USD& JPY/USD\\
	&    & GBP/USD   &AUD/USD\\	& & &NZD/USD\\\hline\hline
	Commodities& Crude Oil  &Gold &Soybeans\\
	& Natural Gas  & Copper &Live Cattle\\	\hline\hline
	Com.-Sectors:  & Energy  &Metals   &Agriculture\\	\hline
\end{tabular}\\

For all contracts, we consider 30 years of daily price data, covering the period from Jan 1, 1990, to Dec 31, 2019. The first two years 1990 and 1991 are merely used to compute the trend strengths at the beginning of 1992 (see below), so the actual regression analysis covers only 28 years. Daily prices $P_i(t)$ were taken from Bloomberg, where $i$ labels the asset and futures are rolled 5 days prior to first notice. We define normalized daily log-returns $R_i(t)$: 
\begin{equation}
R_i(t)={r_i(t)\over \sigma_i}\ ,\ \ \ r_i(t)=\ln {P_i(t)\over P_i(t-1)}\ ,\ \ \ \sigma_i^2=\text{var}(r_i)\ ,\ \ \ \mu_i=\text{mean}(r_i)\ ,\label{return}
\end{equation}
where the long-term daily risk premium $\mu_i$ and the long-term daily standard deviation $\sigma_i$ of a market $i$ are measured over the whole 30-year period.
For some futures markets, the log-returns $r_i(t)$ had to be backtracked or proxied as follows: 

\begin{enumerate}
\item TSE 60 futures: their history begins on Sep 9, 1999. Before, the TSE 60 futures returns are proxied by the S\&P 500 futures returns, which, in our analysis, thus have double weight during that period.
\item Hang Seng index futures: their history begins on Apr 2, 1992. Before, their returns are proxied by Nikkei 225 futures returns. As the regression analysis begins only on Jan 1, 1992, this is a minor data correction.
\item DAX futures: their history begins on Nov 26, 1990. Before, FTSE 100 futures returns are used as a proxy. This data correction is also minor: it merely slightly affects the initial trend strengths at the beginning of 1992, when the analysis begins. 
\item EUR futures: their history begins on May 21, 1998. Before, the Deutsche Mark is used as a substitute for the Euro. We have reconstructed Deutsche Mark futures returns from the spot exchange rate and German/U.S. Libor differentials.  
\item NZD futures: their history begins on May 9, 1997. Before, we have reconstructed the futures returns from the spot exchange rate and the NZD/USD Libor differentials. 
\item German 10-year “Bund” futures: their history begins on Nov 27, 1990. Before, we have reconstructed futures returns from daily German 10-year and short-term interest rates, assuming a duration of $8$. This data correction is also minor: it merely slightly affects the initial trend strengths at the beginning of 2002, when the analysis begins. 
\item Natural gas futures: their history begins on Apr 5, 1990. Before, 1.5-fold levered crude oil futures returns are used as proxies for the natural gas futures returns, using the U.S. Libor rate as the cost of leverage. The 1.5-fold leverage reflects the higher volatility of natural gas compared with crude oil. Again, this data correction is minor, as it merely affects the initial trend strengths at the beginning of 1992, when the analysis starts. 
\end{enumerate}

\subsection{Time Scales}

We will examine the interplay between trends and reversion at 10 different time scales:
\begin{equation}
T_k=2^k\ \text{business days with}\  k\in \{1,2,3,...,10\}\notag
\end{equation}
This represents periods of approximately 2 days, 4 days, 8 days, 3 weeks, 6 weeks, 3 months, 6 months, 1 year, 2 years, and 4 years. Thus, there are 10 different trend strengths at each point in time. A given asset may well be, e.g., in a long-term up-trend at the 1-year time scale, and at the same time in a short-term down-trend at the 3-week time scale.

\subsection{Trend Strengths}

As reviewed in \cite{lev,bru}, there are many different definitions of the strength of a trend, most of which are highly correlated. For the purpose of this study, we need a definition that
\begin{itemize} \addtolength{\itemsep}{-5 pt} 
\item has only a single free parameter, the horizon $T$ (to avoid overfitting historical data)
\item can be computed recursively (which will later help to relate it to critical phenomena)
\end{itemize}

Let us develop the most convenient such definition step by step. For a given time horizon $T$, we define the trend strength $\phi_{i,T}(t)$ of a market $i$ at the end of day $t\in Z$ as a weighted average of past daily returns of that market (i.e., on or before day $t$) - more precisely, of the normalized past daily log-returns (\ref{return}) in excess of the long-term risk premium:
\begin{equation}
\phi_{i,T}(t)=\sum_{n=0}^\infty w_T(n)\cdot \hat R_i(t-n)\ \ \ \text{with}\ \ \ \hat R_i(t-n)=R_i(t-n)-{\mu_i\over\sigma_i},\label{phi}
\end{equation}
where $w_T(n)$ is a weight function for the time scale $T$. Removing the long-term risk premia $\mu_i$ in (\ref{phi}) is necessary to ensure that the long-term expectation value of the trend strengths $\phi_{i,T}$ is zero. If we did not remove the risk premia, very long-term trends in equity- and bond markets, where such risk premia are generally assumed, would almost always be positive and never revert. This mix-up of trends with risk premia would distort our results, as discussed in appendix A2. Note that the long-term risk premia are estimated over the whole time period in (\ref{phi}). However, to avoid any biases, in the out-of-sample cross-validation of section 4 we estimate the risk premia only from the training samples, excluding the validation samples.\\

 We also normalize the weight function $w_T(n)$ such that the trend strength $\phi_{i,T}$ has standard deviation 1. Assuming that market returns on different days are independent from each other (which is true to high accuracy), this implies: 
\begin{equation}
\sum_{n=0}^\infty w_T^2(n)=1.\label{ssq}
\end{equation}
With this normalization, $\phi_{i,T}$ can be regarded as the statistical significance of the trend. E.g., $\phi_{i,T}=2$ represents a highly significant up-trend, while $\phi_{i,T}=-0.5$ represents a weakly significant down-trend. This normalization makes all trend strengths comparable with each other, and will thus allow us to aggregate across different markets and time scales below.\\

The simplest weight function is a {\it step function} (fig. 1, dotted line). In this case, the trend strength $\phi_{i,T}$ is just proportional to the average log-return over the past $T$ days. Unfortunately, this weight function leads to artificial jumps of the trend strength $\phi_{i,T}$ on days when nothing happens, except that an outlier return leaves the rolling time window. \\

This can be avoided by an {\it exponentially decaying} weight function $\tilde w_T(n)$ (fig.1, dashed line). Moreovoer, the corresponding trend strength $\psi$ can now be computed recursively:
\begin{eqnarray}
\tilde w_T(n)&=&M_T\ e^{-2n/T} \ \ \ \text{with normalization factor} \ \ M_T=\sqrt{1-e^{-4/T}},\label{ewma}\\
\psi_{i,T}(t)&=&\sum_{n=0}^\infty\tilde w_T(n)\cdot \hat R_i(t-n)= e^{-2/T} \psi_{i,T}(t-1)+M_T\cdot \hat R_i(t).\label{psi}
\end{eqnarray}

It can be verified that $\tilde w_T$ satisfies  (\ref{ssq}). However, $\psi$ is quite volatile and jumps when an outlier return enters the rolling time  window.

One way to solve this  problem is to use the common definition of $\phi_{i,T}$ in terms of a {\it moving average crossover}: one subtracts the average log-price of asset $i$ over a longer time period $L$ from the average log price of the same asset over a shorter time period $S$. As pointed out in \cite{lev}, this corresponds to a wedge-like weight function (fig. 1, solid line). It makes the trend strength less volatile, as outlier returns affect it only gradually over the time period $S$. It also filters out short-term trends on time scales smaller than $S$, which helps to seperate trends at different time scales from each other.\\

\vspace{1pt}

\begin{figure}[h]\centering
	\includegraphics[height=4.5cm]{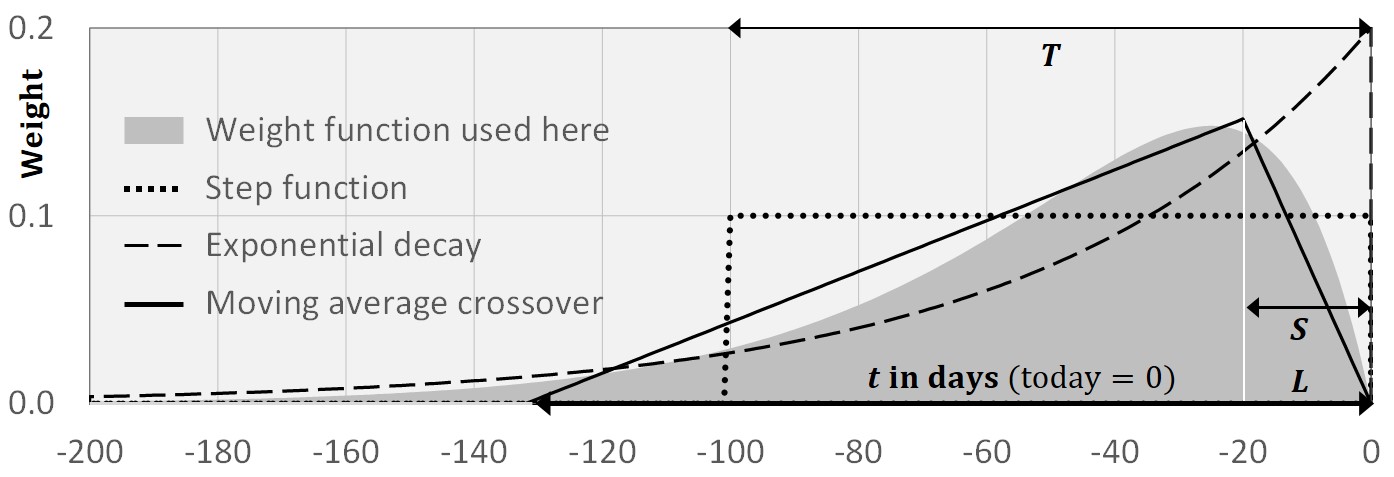} 
	\caption{Our trend strength is defined as a weighted sum of past log-returns. The grey area shows the weight function used in this paper, compared with three standard alternatives. All four weight functions shown here have the same average lookback period.}
	\label{figA}
\end{figure}

Unfortunately, the moving price average has {\it two} parameters $L,S$ (instead of just {\it one} parameter $T$) that must be fitted to the data in any analysis, which tends to reduce the statistical significance of the results. In this article, we will therefore use another similar weight function that involves only the single parameter $T$ (fig. 1, grey area; for comparability with the other weight functions, the figure shows $w_{T/2}$ instead of $w_T$):
\begin{equation}
w_T(n) = N_T\cdot (n+1)\cdot \exp(-{2n\over T})\ \ \ \text{with}\ \ \ N_T={(1-e^{-4/T})^2\over\sqrt{1-e^{-8/T}}}.\label{xex}
\end{equation}
With this normalization factor $N_T$, one can verify that (\ref{ssq}) is indeed satisfied. Moreover, this definition, together with (\ref{psi}), allows for a recursive combined computation of the two variants $\psi,\phi$ of the trend strength (which will be important in section 5): 
\begin{eqnarray}
\phi_{i,T}(t)&=&\sum_{n=0}^\infty w_T(n)\cdot \hat R_i(t-n)= e^{-2/T} \phi_{i,T}(t-1)+{N_T\over M_T}\cdot \psi_{i,T}(t)\label{recursive}
\end{eqnarray}
The "average lookback period" of this trend strength, i.e., the expectation value $E[n+1]$ of the number of days we look back (where "today", i.e. $n=0$, counts as a 1-day lookback), is
\begin{equation}
E[n+1]= \sum_{n=0}^\infty (n+1)\cdot w_T(n)\cdot \big[\sum_{n=0}^\infty w_T(n)\big]^{-1}=T.\label{lookback}
\end{equation}

We have verified that, for a given horizon $T$, all definitions of the trend strength in fact yield quite similar results in our regression analysis of section 4, as long as the weight function rises gradually, decays gradually, and the average lookback period is the same. However, we use (\ref{xex},\ref{recursive}) here, as it is the simplest mathematical function that satisfies these criteria, {\it and} has only the single free parameter $T$, {\it and} can be computed recursively. (\ref{xex}) was originally introduced by the author in 2008 at Syndex Capital Management, and has been used to replicate Managed Futures indices as part of a UCITS fund from 2010-2014. \\

To limit the impact of outlier values of $\phi_{i,T}$ on our results, we will cut it off at $\pm 2.5$ in the actual regression analysis of section 4, i.e., we will use the capped and floored version 
\begin{equation}
\phi_{i,T}^{cap}=\min{(2.5,\max{(-2.5,\phi_{i,T})}}.\notag
\end{equation}
According to \cite{hoyle}, the standard practise of the managed futures industry (which focuses on trends, and not reversion) for this threshold is 2.0. We use the slightly higher value of 2.5, because this will allow us to study more precisely the regime where trends revert, yet it will not give excessive weights to outliers. This is supported by Appendix A1, which compares the results of our regression analysis of section 4 for thresholds from 2.0 to 3.0. For thresholds $>2.5$, we get a higher adjusted R-squared. However, the results are less robust, and the statistical significance of the regression betas decreases. For thresholds $<2.5$, the reversion regime would be largely removed from the analysis, leading to a lower adjusted R-squared without improving the overall statistical significance of the results. 

\newpage
\subsection{Database}

Table 2 displays a small extract of the resulting database for our analysis. Only two of the 7305 business days and only three of the 24 markets are shown. The third column shows the normalized daily log-returns (\ref{return}), which have standard deviation 1. The 7305 business days cover only the 28-year period from Jan 1992 - Dec 2019, because the first two years 1990-1992 were only used to compute the initial trend strengths at the beginning of 1992. The full table with $7305\times24=175'320$ lines is published along with this paper.\\

{\small
\begin{tabular}{ |p{0.5cm}|p{1.5cm}|p{0.8cm}||p{0.7cm}|p{0.7cm}|p{0.7cm}|p{0.7cm}|p{0.7cm}|p{0.7cm}|p{0.7cm}|p{0.7cm}|p{0.7cm}|p{0.7cm}|	}\hline
	\multicolumn{3}{|c||}{\it Table 2: Database } &\multicolumn{10}{|c|}{Trend strengths on previous day for 10 time scales}\\	\hline
	Day& Market  &$ R_i(t)$&2d   &4d & 8d & 3w & 6w & 3m & 6m & 1y & 2y & 4y\\	\hline\hline
1	&	S\&P 500	&	-0.2	&	0.3	&	0.7	&	1.0	&	0.6	&	0.2	&	0.3	&	0.6	&	0.6	&	1.0	&	1.6	\\ \hline
1	&	EUR/\$	&	-0.1	&	0.2	&	0.2	&	0.0	&	-0.4	&	-0.6	&	-0.6	&	-0.8	&	-0.9	&	-0.7	&	-0.8	\\ \hline
1	&	Gold	&	-0.5	&	-0.3	&	-0.7	&	-0.7	&	0.1	&	1.1	&	1.5	&	1.1	&	0.4	&	0.1	&	-0.4	\\ \hline\hline
2	&	S\&P 500	&	-0.3	&	-0.1	&	0.4	&	0.9	&	0.7	&	0.2	&	0.3	&	0.6	&	0.6	&	1.0	&	1.6	\\ \hline
2	&	EUR/\$	&	-1.0	&	-0.7	&	-0.2	&	-0.2	&	-0.4	&	-0.6	&	-0.6	&	-0.8	&	-0.9	&	-0.7	&	-0.8	\\ \hline
2	&	Gold	&	0.8	&	0.3	&	-0.2	&	-0.6	&	0.1	&	1.1	&	1.5	&	1.1	&	0.4	&	0.1	&	-0.4	\\ \hline
\end{tabular}}\\

\section{Qualitative Observations}

This section begins with an exploratory analysis of our data. The analysis in this section is only qualitative, but it serves to motivate the specific quantitative, statistically rigorous regression analysis of the following section. We stress again that our aim is {\it not} to improve futures trading strategies, which would have to include risk limits, trading cost minimization, and other features. Rather, we simply want to empirically measure and model the small autocorrelations of market returns as accurately as possible as a basis for future work.

\subsection{Next-day Return vs. Trend Strength}

We use the data of table 2 to measure the expected daily return of a futures market as a function of the trend strengths in that market on the previous day. To this end, we first construct  $7305\cdot 24\cdot 10=1'753'200$ pairs of data. Each pair consists of the normalized log-return $R_{i}(t)$ in that market on day $t$, and one of the 10 trend strengths $\phi_{i,T}(t-1)$ on the previous day. So each return appears in 10 data pairs, each time paired with a different trend strength. \\

We then group those pairs into 15 bins of increasing trend strength from $-\infty$ to -13/6, -13/6 to -11/6, ... ,-1/6 to+1/6, ... ,11/6 to 13/6, 13/6 to $\infty$. The mean trend strength within each bin is shown on the x-axis of fig. 2 (left). Within each bin, we average over the normalized return on the day after the trend has been measured. To obtain statistically significant results, we aggregate over the 28 years of daily returns for each market, across all 24 markets, and across different time scales. Fig. 2 (left) shows the results for the 4 monthly trend strengths (i.e., aggregated over $T=$ 6 weeks, 3 months, 6 months, and 1 year).\

\begin{figure}[h]\centering
	\includegraphics[height=5.7cm]{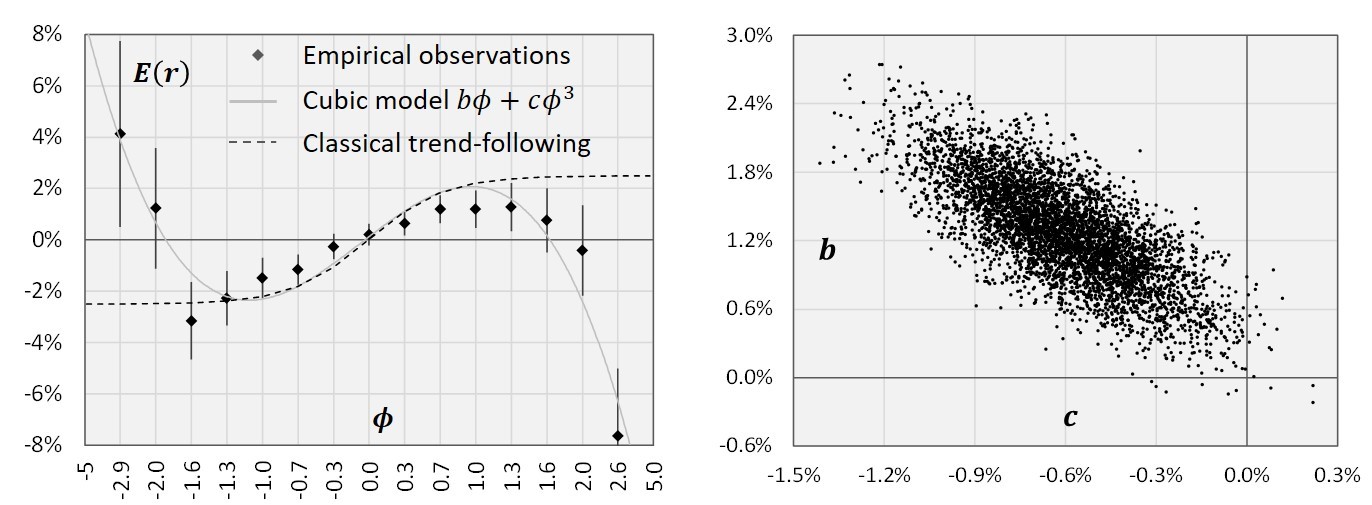} 
	\caption{ {\it Left}: The expectation value $E(r)$ of the next day’s return of a futures market is a nonlinear function of the current trend strength $\phi$. As verified by the extensive statistical analysis of section 4, it can be modeled by a cubic polynomial of $\phi$, whose linear term $b\phi$ (with $b>0$) represents trend-persistence, and whose cubic term $c\phi^3$ (with $c<0$) represents trend-reversion.  
{\it Right}: As confirmed by bootstrapping in section 4, the regression coefficients $b$ and $c$ corresponding to the linear and cubic terms are statistically highly significant.}
	\label{figA}
\end{figure}

We observe that the average next-day return is close to zero at zero trend strength, and grows linearly with the trend strength for small strengths. As the trend strength increases further, the average next-day return peaks, then decreases again until it becomes zero somewhere below trend strength 2. For even stronger trends, the average return decreases dramatically. This behavior is mirrored on the left-hand side of the graph for down-trends. We have verified that this pattern remains almost the same if another day of delay is added, i.e., if the next-day return in our data pairs is replaced by the return 2 days later.\\

Thus, trends tend to revert when they become too strong. This makes sense intuitively: after strong trends, markets tend to be overbought or oversold, so one expects a reversion to "value". Our analysis quantifies where exactly this happens: below a critical trend strength of 2, before trends become strongly statistically significant. Note that this is not in line with classical trend-following, which would follow the trend no matter how strong it becomes. The dashed line in fig. 2 (left) indicates the trading position that a classical trend-follower would take as a function of the trend strength. 

\subsection{Dependence on the Time Scale}

In a next step, we analyze how the pattern observed in the previous sub-section depends on the time scale. To this end, we refine the bins used above: we split each bin into 10 smaller bins, one for each of the 10 time scales. The resulting 15$\times$10 refined bins are now too small and the results too noisy. To reduce the noise, we average the next-day returns over blocks of 3x3 neigboring bins (resp. 3x2 or 2x3 bins at the borders, 2x2 bins at the corners of the matrix of 15$\times$10 bins), weighted by the number of returns in each bin. This yields the heat map of fig. 3 (left). Fig. 2 (left) can be thought of as a horizontal cross-section through this heat map along the dashed line.\\

\begin{figure}[h]\centering
	\includegraphics[height=5.7cm]{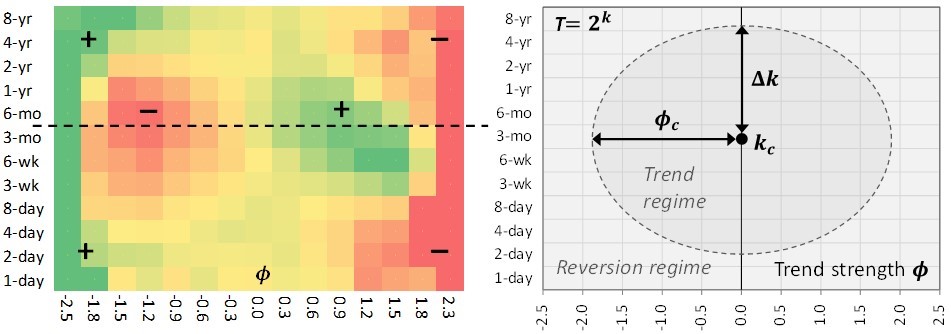} 
	\caption{ {\it Left}: A heat map shows how the expectation value of tomorrow’s return depends both on today’s trend strength $\phi$ and its time horizon. Fig. 2 (left) can be thought of as a cross-section of Fig. 3 (left) along the dashed line. {\it Right}: The polynomial regression analysis of section 4 models the pattern of Fig. 3 (left) by an elliptic regime within which trends are persistent, and outside of which they revert. The values of the center and of the semi-axes of the ellipse are statistically highly significant, as confirmed by bootstrapping.}
	\label{figA}
\end{figure}

We observe that the trend- and reversion pattern of fig. 2 (left) is strongest on time scales from 1 month to 1 year. This is in line with the fact that typical trend-followers operate on those time scales. As the time scale increases or decreases, the pattern becomes weaker. The region where markets trend seems to disappear both for time scales of the order of economic cycles (several years) and for intra-week time scales. The phenomenon of reversion, on the other hand, appears to remain strong at all time scales.

\subsection{Counting Degrees of Freedom}

As emphasized in \cite{har}, one must be very conservative in introducing new factors and parameters in financial market models. Before modeling the observed patterns in detail, let us therefore do a back-of-the-envelope calculation of how many parameters we can hope to fit in our model without over-fitting our daily return data, and what fraction of the variance of these returns we can hope to explain by trend factors.\\

Our $7'305\cdot 24=175'320$ daily log-returns are not independent, because the 24 markets are correlated with each other. How many independent markets are there? The daily returns are normalized to have variance 1. For a portfolio that invests $1/24$ in each market, we find a variance of $\sigma^2\sim 1/8$, just as if it contained $n_m=8$ independent assets. A principal component analysis confirms that the first 8 (resp. 12) principal components explain 65\% (resp. 80\%) of the variance of the returns of our 24 markets. In this sense, these returns effectively live in a space of dimension $n_m\sim8$. Adding more markets to our 24 time series does not significantly increase $n_m$.\\

What is the highest annualized Sharpe ratio $S$ that one can hope to achieve by systematically trading a broadly diversified set of highly liquid futures markets based on trends and reversion? Experience with the Managed Futures ("CTA") industry suggests that $S$ can be at best 1. The small number of CTA's that have achieved a higher Sharpe ratio for several years in a row presumably also pursue other strategies that are not purely based on trends, or they are not market-neutral (in the sense of zero net exposure to each market over time).\\

An annualized Sharpe ratio of $S=1$ implies a daily Sharpe ratio $\rho$ for each market of
\begin{equation}
\rho={S\over \sqrt{260\cdot n_m}}\sim 0.02.\notag
\end{equation}
So the predicted next-day return of a market has a correlation of $\rho=0.02$ with the actual next-day return. E.g., if we only try to predict the sign of the next return, we can at best hope to be right on 51 and wrong on 49 out of 100 days. The adjusted R-squared (achieved out-of-sample in real trading) is then $R^2_{adj}\sim \rho^2=4$ basis points ($1 bp=4\cdot 10^{-4}$). Clearly, the variance of financial market returns is overwhelmingly due to random noise.\\

If we fit $k$ parameters to our data, and if our returns were independent and identically distributed ("iid"), then, for small $R^2$, the adjusted R-squared would be approximately
\begin{equation}
R^2_{adj}\sim R^2-{k\over N}\sim 4\ bp\ \ \ \text{with}\ \ \ N=260\cdot n_m\cdot Y\ \text{data points} \label{envelope}
\end{equation}
 for $Y$ years of daily data. If we require that the correction for the in-sample bias does not erode more than 20\% of our $R^2$, then we conclude that we cannot fit more than $k\sim N\cdot 1\ bp\sim 6$ parameters to our 28 years of data. This 20\%-requirement is not too conservative, as our returns are only approximately "iid", and therefore the actual correction per fitted parameter will be higher than 20\% (below, cross-validation will show that it is indeed more than twice as big). We conclude that we must use parameters wisely, and not "waste" them on features that may be artifacts of our limited data set. 

\section{Regression Analysis}

In this section, we confirm and quantify the observations of the previous section by nonlinear regression based on ordinary least squares. To this end, we model the next-day return of a market as a polynomial function of {\it both} the current trend strength in that market {\it and} its time scale. This regression is performed directly on the underlying 1'753'200 pairs of data, not on the bins we have defined in the previous section. Thus, our results are independent of any choice of how to split the data into bins.

\subsection{Dependence on the Trend Strength}

The graph in fig. 2 (left) suggests to model the next-day normalized log-return $R(t+1)$ (\ref{return}) as a polynomial of the current trend strength $\phi(t)$ (\ref{phi}) across all markets and time scales:
\begin{equation}
R(t+1)=a+b\cdot \phi(t)+d\cdot \phi^2(t)+c\cdot \phi^3(t)+...+\epsilon(t+1),\label{cubic}
\end{equation}
where $\epsilon$ represents random noise, and $a$ measures the average risk premium $\mu_i/\sigma_i$ across all assets. Similar models with a polynomial random force have been postulated previously, notably by econophysicists with a background in critical phenomena \cite{bouchard, ide, puk,puk1}. Our observations of the previous section give clear empirical support for a polynomial ansatz. We will discuss the relationship with critical phenomena in more detail in section 5.\\ \\

\begin{tabular}{ |p{2.5cm}||p{3.5cm}|p{2.5cm}|p{2.5cm}|  }	\hline
 	{\it Table 3} & \multicolumn{3}{|l|}{Regression with linear and cubic terms} \\ \hline
	{Coefficient}& Value  &Error &t-statistics\\	\hline
	$a$   & $1.33$\%    &$\pm 0.41$\% &  $3.3$\\  
	$b$ &$1.29$\% & $\pm 0.43$\% &  $3.0$ \\	
	$c$ &  $-0.62$\%  &  $\pm 0.23$\% & $2.7$\\ \hline\hline
	{R-squared} & Single time scales & \multicolumn{2}{|l|}{Aggregated across time scales} \\ \hline
	$R^2$ &$1.31$\ bp & \multicolumn{2}{|l|}{$4.91$\ bp}  \\	
	$R^2_{adj}$ &  $1.03$\ bp  &  \multicolumn{2}{|l|}{$3.98$\ bp}  \\ \hline
\end{tabular}\\

We have performed a corresponding regression analysis on the 1'753'200 data pairs $\{r_{t+1},\phi_t\}$. Using only the linear and cubic terms of (\ref{cubic}) yields the results of table 3. \\

Since market returns cannot be assumed to be independent, identically distributed normal variables, we cannot trust the usual estimates of the t-statistics, adjusted R-squared, and F-statistic. Instead, the test statistics shown in table 3 are measured empirically as follows:
\begin{itemize}
\item  The standard errors of the coefficients and their t-statistics are computed by bootstrapping: from the 7305 days, we randomly create 5000 new samples of 7305 days each, with replacement. I.e., some days occur several times in a new sample, while other days do not occur at all. Regression on each new sample of days yields the distribution of 5000 regression coefficients $b$ and $c$ shown in fig. 2 (right). The errors of the coefficients in table 3 represent half the difference between the 84th and 16th percentile, which equals the standard deviation in the case of a normal distribution.
\item The adjusted R-squared is computed by 15-fold cross validation: we split our 7305-day time window into 15 sub-windows of 487 consecutive days each. For each sub-window, we predict the next-day returns based on the betas obtained by regression on the other 14 sub-windows. The square of the correlation between the predicted and the actual returns is the out-of-sample $R$-squared $R^2_{adj}$ reported in table 3. 
\item  
Table 3 also reports ${R}^2$ and ${R}^2_{adj}$ "aggregated across time scales". Those are based on using the equally-weighted mean of the 10 trend strengths on each day to predict the next-day return for each market. I.e., we combine the 10 different trend factors into a single one, which naturally has a higher predictive power than each single factor by itself. This regression is thus performed on only $7305\cdot24=175'320$ pairs of data. 
\item
The $F$-statistics can be computed numerically to be $F=4.6$ with a $p$-Value of $0.7\%$ by modelling the distribution of regression coefficients in fig. 2 (right) by an elliptical distribution. However, the distributions in subsequent sections are not even approximately elliptical. We will therefore use $R_{adj}^2$ and not $F$ to compare the out-of-sample explanatory power of our models with each other.  
\end{itemize}

The regression results of table 3 confirm and quantify our conclusions from the previous section. We see that the values of $b$ and $c$ - although very small - are statistically highly significant, despite the fact that market returns are neither normally distributed, nor independent, nor identically distributed. So is the average long-term risk premium $a$. The overall result is significant at the 99\% level. The aggregated out-of-sample ${R}^2_{adj}$ that combines the predictions from all 10 time scales matches our initial expectation of $4\ bp$ (\ref{envelope}). Note that the correction for the aggregated in-sample bias, ${R}^2-{R}^2_{adj}=0.93\ bp$, is much bigger than what would have been expected if returns were "iid", namely $2/(260\cdot8\cdot28)=0.34\ bp$.\\

We have also tested the quadratic, quartic and quintic terms in $\phi_T$ in (\ref{cubic}). None of them turned out to be statistically significant at the 95\% level. We therefore drop them from our analysis to avoid over-fitting the historical data (i.p., the t-statistics for $d$ is below 1).

\subsection{Dependence on the Time Scale}

Next, we try to refine our model by measuring the dependence of the coefficients b and c on the time scale, the asset class, and the time period. We begin with the time scale $T$: we model the expected return as a function of both the trend strength {\it and} $T$, trying to replicate fig. 3 (left). We first repeat the linear and cubic regression (\ref{cubic}) of the previous sub-section for each of the 10 time scales seperately. The resulting coefficients $b(T)$ and $c(T)$ are plotted in fig. 4 (we neglect the overall risk premium $a$, which is not the focus of this paper).\\

From the coefficient $b$ of the linear term, which models trends, we observe that trend-following works best at time scales from 3 months to 1 year, where $b$ peaks. This appears to be in line with the time scales on which typical CTAs follow trends. Even at those scales, the critical trend strength
\begin{equation}
\phi_c=\sqrt{-b/c}\ \le1.91\ ,\notag
\end{equation}
beyond which trends tend to revert, is below 2. So trends never become strongly significant. For scales below a few days and above several years, $b$ seems to go to zero, which means that trends are not persistent there. This is consistent with the heat map in fig. 2 (right).

\begin{figure}[h]\centering
	\includegraphics[height=5.5cm]{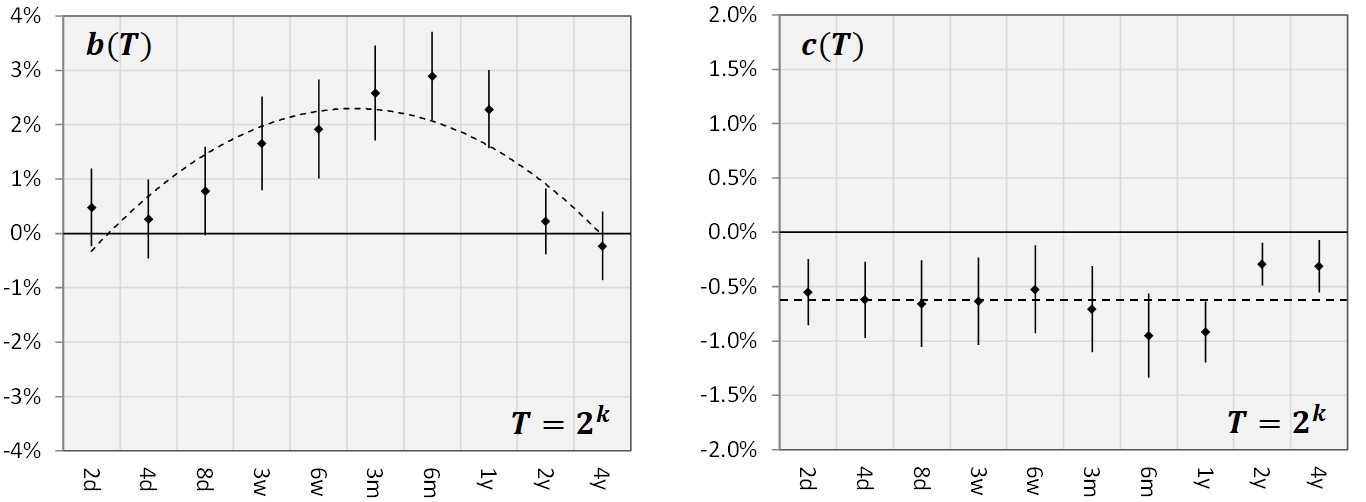} 
	\caption{{\it Left:} The coefficient $b(T)$ of the linear term (corresponding to the trending of markets) peaks at time scales $T$ of 3 months to 1 year. Its scale dependence is modeled by a parabola. {\it Right:} The coefficient $c$  of the cubic term (corresponding to the reversion of markets) does not show a clear dependence on the time scale.}
	\label{figA}
\end{figure}

On the other hand, the coefficient $c$ of the cubic term, which ensures that trends revert, is quite stable, except that its magnitude appears to be somewhat lower for the 2- and 4-year scales. The 2- and 4-year results must be taken with a grain of salt, though, as there are only 14 independent 2-year trends and 7 independent 4-year trends in our 28-year time window. Indeed, a preliminary check based on 60 years of monthly returns resulted in $c\sim-0.6\%$ at the 8-year scale. The available data thus indicate that, unlike trend-following, mean reversion works at all time scales. This is also consistent with our earlier observation from the heat map in fig. 3 (left).\\

To quantify these observations, we refine our regression ansatz (\ref{cubic}). We continue to model the cubic coefficient $c$ by a constant, but we model the dependence of the linear coefficient $b(k)$ on the logarithm $k$ of the time scale $T=2^k$ by a parabola:
\begin{eqnarray}
b(k)&=& b-e\cdot (k-k_0)^2\  \notag\\ 
\Rightarrow R(t+1)&=&b\cdot\Big\{1-{(k-k_0)^2\over(\Delta k)^2}\Big\}\cdot \phi(t)+c\cdot \phi^3(t)+\epsilon(t+1)\  \label{scale}
\end{eqnarray}
with $(\Delta k)^2={b/e}$. The critical trend strength $\phi_c(k)=({-b(k)/c})^{1/2}$, at which the expected return $E(R_{t+1})$ is zero (without the noise $\epsilon$), and beyond which trends revert, is then an ellipse with semi-axes $\Delta k$ and $\phi_c(k_0)$. Altogether, we now fit 4 parameters to our data:

\begin{itemize}\addtolength{\itemsep}{-6 pt} 
\item
The "persistence of trends" $b$, i.e. the value of $b(k)$ at its peak
\item
The "strength of reversion" $c$ 
\item
The range $k_0\pm \Delta k$ of the log of the time scales $T=2^k$ at which markets may trend.
\end{itemize}\vspace{10pt}

\begin{tabular}{ |p{2.5cm}||p{3.5cm}|p{2.5cm}|p{2.5cm}|  }	\hline
 	{\it Table 4} & \multicolumn{3}{|l|}{Refined regression with 4 parameters} \\ \hline
	{Coefficient}& Value  &Error &t-Stat.\\	\hline
	$b$ &$2.00$\% & $\pm 0.48$\% &  $4.2$ \\	
	$c$ &  $-0.63$\%  &  $\pm 0.24$\% & $2.6$\\ 
	$k_0$ &  $5.78$  &  $\pm 0.67$ & 8.6 \\ 
	$\Delta k$ &  $4.87$  &  $\pm 1.09$ & $4.5$\\ \hline\hline
	{R-squared} & Single time scales & \multicolumn{2}{|l|}{Aggregated across time scales} \\ \hline
	$R^2$ &$1.64$\ bp & \multicolumn{2}{|l|}{$7.51$\ bp}  \\	
	$R^2_{adj}$ &  $1.22$\ bp  &  \multicolumn{2}{|l|}{$6.04$\ bp}  \\ \hline
\end{tabular}\\ \\

A nonlinear regression on the full underlying data set yields the results of table 4. Fig. 3 (right) plots the elliptic region, which seperates the “trend regime” (inside) from the “reversion regime” (outside). For its second semi-axis, we find $\phi_c(k_0)=1.78\pm0.32$. This quantifies the empirical heat map in fig. 3 (left) and confirms that highly significant trends of strength (i.e., $t$-statistics) $\phi_c\ge2$ always tend to revert. \\

The errors of the regression parameters in table 4 are again computed by bootstrapping. The distribution of $b$ and $c$ looks the same as in the univariate case (Fig. 2, right). Fig 5 (left) plots the distribution of the values of the center $k_0$ and the semi-axis $\Delta k$ of the ellipse that separates the trending regime from the reversion regime. The "aggregate" $R^2$ and $R^2_{adj}$ in table 4 now refer to a single factor that is a linear combination of the 10 trend strengths for the 10 time scales, weighted by a parabolic weight function proportional to $b(k)$. Note that the aggregated adjusted R-squared now exceeds our original expectation (\ref{envelope}) of $4\ bp$.

\begin{figure}[h]\centering
	\includegraphics[height=5.7cm]{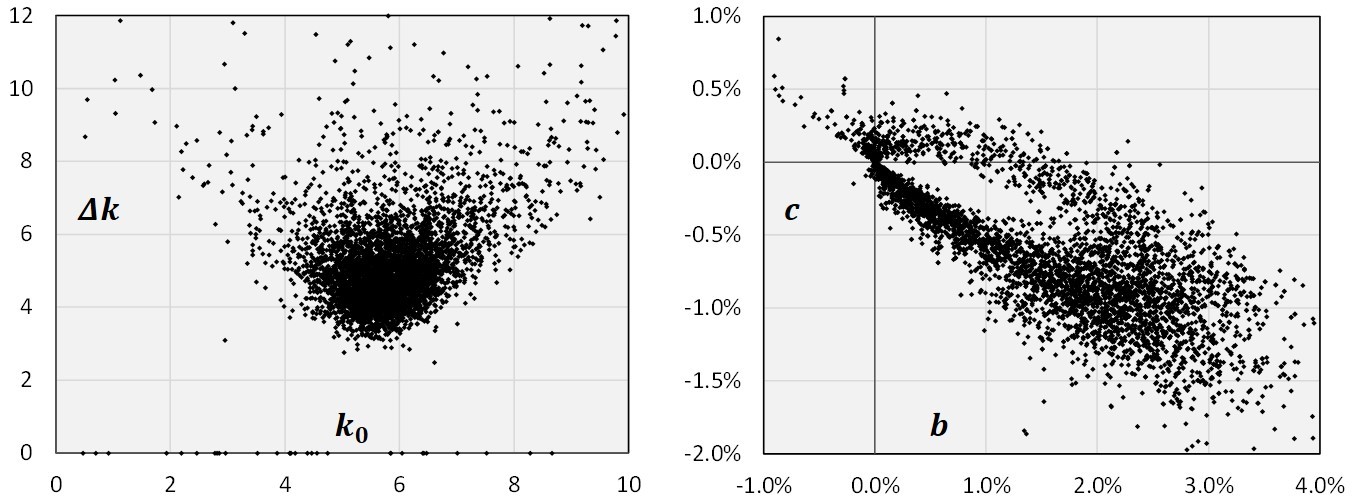} 
	\caption{{\it Left:} Distribution of the regression coefficients for the center $k_0$ and width $\Delta k$ of the elliptic region within which trends are persistent, as obtained by bootstrapping. {\it Right}: Distribution of the linear and cubic regression coefficients $b,c$ for a rejected alternative model.}
	\label{figA}
\end{figure}

We have tried to further refine ansatz (\ref{scale}). First, $b(k)$ in fig. 4 (left) seems to be tilted to the right, which could be accounted for by models such as $b(k)\sim  b-e\cdot (k-k_0)^2+f\cdot(k-k_0)^3$, or $b(k)\sim \exp(f\cdot k)\cos((k-k_0)/\Delta k)$. We find that such models increase the adjusted $R$-squared at best marginally. Therefore, we use the simplest model (\ref{scale}) in this paper, to avoid over-fitting the historical data. \\

Second, we also tested for a polynomial dependence of $c$ on $k$. The most significant ansatz was that $c(k)$ is also a parabola proportional to $-b(k)$. In this case, the critical trend strength is constant across all time scales, and the region within which markets trend is rectangular instead of elliptic. The distribution of the parameters $b(k_0),c(k_0)$ then turns out to have the shape of the stretched annulus shown in fig. 5 (right). However, this scenario seems less likely, as it yields a much lower adjusted R-squared (0.77 bp).

\subsection{Dependence on the Asset Class}

Can we refine our 4-parameter-model further by distingushing between asset classes, i.e., by fitting seperate regression parameters for equities, bonds, currencies and commodities? \\

To test this, we have repeated the regression analysis of the previous section for these 4 sub-sets of our data. Fig. 6 (left) shows the 16th, 50th and 84th percentile of the values of the 4 regression parameters for each asset class, divided by the values of the regression parameters for the overall sample. E.g., for equities, the quantiles for $b$ are  $(1.80\%, 2.82\%, 4.01\%)$, which are multiples of $(0.90,1.41,2.01)$ of the overall regression coefficient $2.00\%$ (see table 3). Those multiples are what is shown in the first bar of fig. 6 (left).\\

\begin{figure}[h]\centering
	\includegraphics[height=5.7cm]{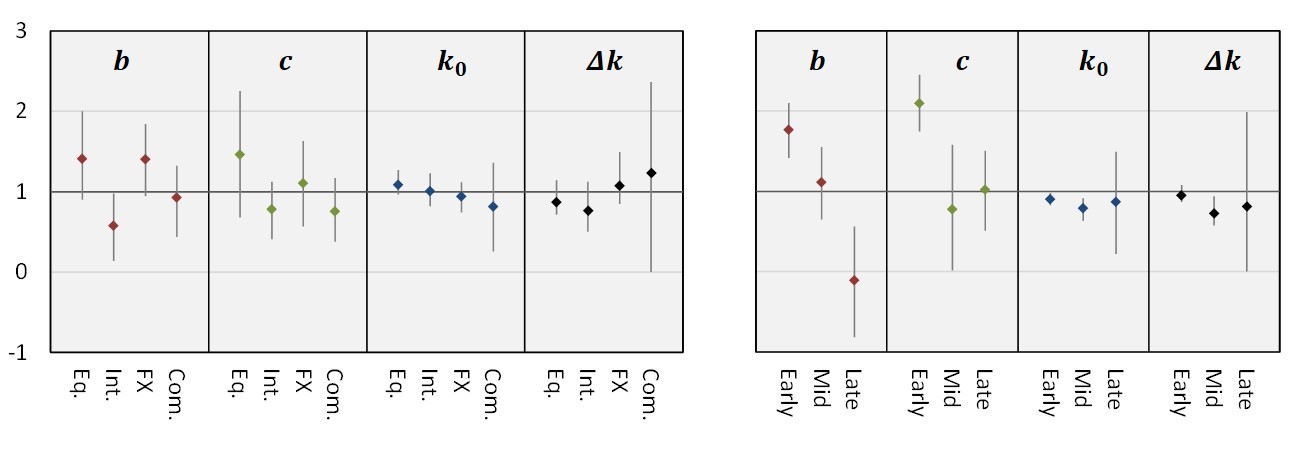} 
	\caption{{\it Left: } Ratios of the values of our 4 regression parameters for equities, interest rates, FX rates and commodities, divided by their overall values across all asset classes. The ratios do not differ significantly from 1. {\it Right:} The analoguous ratios for the early, middle and late third of the time period. At least $b$ has decreased significantly over time.}
	\label{N}
\end{figure}

For each asset class, we observe that the values of all four parameters are within one standard error of the overall parameter values. Thus, based on our data set, we cannot justify fitting different parameters of our trend-reversion model to individual asset classes, let alone to individual assets. This is consistent with our back-of-the-envelope caculation of subsection 3.4, which suggests that we cannot fit as many as $4\times4=16$ parameters for the four different asset classes to our data. We thus use a single, universal model for all assets.

\subsection{Dependence on the Time Period}

Lastly, we investigate how our patterns have been evolving in time. First, we split up the 28-year (7305-day) time window into three non-overlapping sub-periods of 2435 days each:
An early period from Jan 1992 to Apr 2001, a middle period from May 2001 to Aug 2010, and a late period from Sep 2010 to Dec 2019.\\

Fig. 6 (right) shows the 16th, 50th and 84th percentile of the values of the 4 regression parameters for each of these time-windows, again divided by the values of the regression parameters for the overall sample. The persistence of trends $b$ has consistently and significantly decreased over time. With less consistency, this can also be observed for strength of reversion $c$, while no clear trend is visible for the range $k_0\pm\Delta k$, within which trends persist. The decrease of $b$ is in line with the industry observation that trend-following no longer works as well as it used to: markets seem to have become more efficient in this respect. Given the decrease in trading costs, an increase in algorithmic trading, and an increase in assets under management invested in trend-following, this is not surprising.\\

To verify and quantify these observations, let us introduce time $t$, measured in years, with its origin $t=0$ on Dec 31, 2005, the center of our 28-year time window. We now further refine our model (\ref{scale}) by including linear trends in $b$ and $c$ while leaving $k_0$ and $\Delta k$ constant:
\begin{equation}
b(t) = \bar b\cdot(1-Q_b\cdot t)\ \ \ ,\ \ \ c(t) = \bar c\cdot(1-Q_c\cdot t)\notag
\end{equation}
The results of a regression analysis, including bootstrapping and cross-validation, are shown in table 5. The decrease of $c$, which measures the strength of reversion, is only weakly significant. However, the decrease of $b$, which measures the persistence of trends, is significant at the 97.5\% confidence level. In principle, we could compute the year $Y_0$, in which $b(t)=0$:
\begin{equation}
Y_0=2005+{1\over Q_b}\ \in\ \{2012, 2028\}\ \ \ \text{with expectation value}\ \ \ Y_0\sim2017.\notag
\end{equation}
Thus, if one were to take this linear down-trend of $b$ literally, one would conclude that the phenomenon of persistent market trends may have already disappeared. However, there are other scenarios for the time decay of the persistence of trends that are consistent with our data. E.g., for an exponential decay scenario, in which trends never disappear, we find an only slightly lower adjusted R-squared of $1.39\ bp$ instead of $1.52\ bp$, with
\begin{equation}
b\sim \bar b\cdot e^{-Qt}\ \ \ \text{with decay rate}\ \ \ Q\sim (24\ \text{years})^{-1}.\notag
\end{equation}\\
\begin{tabular}{ |p{2.5cm}||p{3.5cm}|p{2.5cm}|p{2.5cm}|  }	\hline
 	{\it Table 5} & \multicolumn{3}{|l|}{Refined regression with 6 parameters} \\ \hline
	{Coefficient}& Value  &Error &t-Stat.\\	\hline
	$\bar b$ & $1.91$\% & $\pm 0.49$\% &  $3.9$ \\	
	$\bar c$ &  $-0.62$\% & $\pm 0.25$\% &  $2.5$\\ 
	$k_0$ &  $5.83$ & $\pm 0.50$ &  $11.7$\\ 
	$\Delta k$ &  $4.97$ & $\pm 0.69$ &  $7.2$\\ 
	$Q_b$ &  0.088 & $\pm$ 0.045 &  $2.0$ \\
	$Q_c$ &  0.047 & $\pm$ 0.052  &  $0.9$ \\ \hline\hline
	{R-squared} & Single time scales & \multicolumn{2}{|l|}{Aggregated across time scales} \\ \hline
	$R^2$  & $1.97$\ bp & \multicolumn{2}{|c|}{$8.90$\ bp}   \\	
	$R^2_{adj}$ &  $1.49$\ bp & \multicolumn{2}{|c|}{$6.98$\ bp}   \\ \hline
\end{tabular}\\  \\

We have also tested scenarios where all 4 parameters or other subsets of them change at different rates, but found that all of these scenarios significantly reduce the adjusted R-squared. It is left for future work to investigate the time evolution of the pattern of trends and reversion in more detail.

\section{Analogies with Critical Phenomena}

In this section, we point out some striking analogies between the empirical observations of sections 3 and 4 and critical phenomena in statistical mechanics. Analogies between financial markets and critical phenomena, such as scaling relations, have long been observed \cite{mant}. Our results go further: they seem to directly and specifically identify the trend strength with the order parameter of a Landau-type mean field theory with a quartic potential.\\ 

Analogies with critical phenomena are plausible, if financial markets are regarded as statistical mechanical systems, whose microscopic constituents are the Buy/Sell orders of individual traders. It is conceivable that these orders can be modeled by degrees of freedom that sit on the vertices of a hypothetical "social network of traders". These degrees of freedom may interact with each other in analogy with spins on a lattice, thereby creating the macroscopic phenomena of trends (herding behavior) and reversion (contrarian behavior). To imitate these phenomena and their interplay, various spin- and agent models have been proposed in the literature (see, e.g., \cite{lux,farmer}, and \cite{sornette} for a recent review).\\

Candidates for the "social network of traders" include small-world networks \cite{watts}, scale-free networks \cite{barbasi}, or the Feynman diagrams of large-N field theory \cite{brezin}. For a recent review of candidates for social networks, see \cite{cimi}. To our knowledge, no convincing specific model has emerged as a consensus so far. Our results provide an empirical basis for accepting or rejecting such candidates: any statistical-mechanical model of financial markets, if accurate, must replicate the interplay of trends and reversion observed in this paper.\\

To make this precise, let us first reap the benefits of our recursive definitions (\ref{psi},\ref{recursive}) of the trend strength, which lead to simple differential equations in the "continuum limit" $T\gg1$:
\begin{equation}
({d\over dt}+{2\over T})\ \psi(t)={2\over\sqrt{T}}\cdot \hat R_i(t), \ \ \ ({d\over dt}+{2\over T})\ \phi(t)={2\sqrt{2}\over T}\ \psi(t).\label{contin}
\end{equation}
To be specific, let us focus on the 6-month time horizon, i.e., $T=2^7=128$ trading days (the results for other horizons are similar). Combining (\ref{contin}) with the ansatz (\ref{cubic}) implies the following second-order stochastic differential equation for the trend strength $\phi$:\vspace{5pt}
\begin{equation}
({d\over dt}+{1\over 64})^2 \phi(t)=-{\partial\over\partial\phi}V(\phi)+{1\over 256}\epsilon(t)\ \ \ \text{with}\ \ \ V(\phi)=-{b\over 512}\cdot \phi^2+{\vert c\vert\over 1024} \cdot \phi^4,\label{stoch}\vspace{5pt}
\end{equation}
with rescaled random noise $\epsilon$. Its simpler cousin $\psi$ in (\ref{psi}) obeys a first-order equation:
\begin{equation}
({d\over dt}+{1\over 64})\psi(t)=-{\partial\over\partial\phi}\tilde V(\psi)+{1\over4\sqrt{2}}\epsilon(t)\ \ \ \text{with}\ \ \ \tilde V(\psi)=-{\tilde b\over4}\cdot \psi^2+{\vert \tilde c\vert\over8} \cdot \psi^4,\label{lange}
\end{equation}
with the following empirical parameter values, as measured by a regression analysis that is analogous to that reported in section 4 for $\psi$: 
$$b=2.94\%,\ c=-0.95\%,\ \tilde b=1.79\%,\ \tilde c=-0.66\%.$$
(\ref{lange}) is the purely dissipative Langevin equation, which is reminiscent of the earlier description \cite{bouchard} of the dynamics of financial markets at intraday scales by another Langevin equation. In the theory of critical phenomena, the Langevin equation is well-known to describe the dynamics of the order parameter of certain statistical mechanical systems near second-order phase transitions \cite{hohenberg,ZJ}. This is consistent with the conjecture that the trend strength (defined as either $\phi$ or $\psi$) plays the role of an order parameter, in analogy with the magnetization in spin models.\\

To take the analogy further, statistical mechanical systems near second-order phase transitions are characterized by universal critical exponents. E.g., a scalar field theory with a $\phi^4$ potential similar to the potentials $V$ in (\ref{stoch},\ref{lange}) describes water and steam and other physical systems in the same universality class (such CO$_2$ or the Ising model) near their critical points \cite{ZJ}. For all systems within this universality class, the parameters $b$ and $c$ show the same scaling behavior as a function of the length scale $L$ (e.g., $b\sim L^\kappa$ for some exponent $\kappa$). In critical dynamics, scaling with $L$ also translates into a scaling with the time horizon $T$ \cite{hohenberg}. \\

In section 4, we have seen that - within the limits of statistical significance - the values of the coefficients $b$ and $c$ are the same for very different markets, such as equity indices, bonds, FX-rates, and commodities. The parameters $k_0$ and $\Delta k$ in (\ref{scale}), which characterize how $b$ behaves under a rescaling of the time horizon $T$, are also the same. This could be an expression of universality and scaling in financial markets. To confirm this, it will be key to examine how the scaling behavior in (\ref{scale}) extends to intra-day and multi-year time horizons $T=2^k$ with $k>10$ or $k<1$. For example, it might reflect a complex critical exponent \cite{sor}. Together with the stochastic differential equations (\ref{stoch},\ref{lange}), the empirically observed scaling behavior may uniquely specify a particular social network that models financial markets. \\

To conclude this section, let us compare with some previous work. In \cite{ide}, a related model for the dynamics of asset prices was postulated. The role of the trend was played by the deviation of the current asset price from its unknown "value". Terms of any order were considered in the polynomial potential, and the corresponding classical solutions were discussed. Compared with \cite{ide}, our trends are measurable, and we focus on a quartic potential, empirically observe the values of its coefficients and their scale dependence, and provide a simple and intuitive map between the quadratic (quartic) terms and trends (reversion). \\

In \cite{puk}, another model with a polynomial random force similar to (\ref{cubic}) was postulated. The trend strength was defined by a moving average crossover (which does not lead to exact differential operators such as (\ref{contin}) in the continuum limit). This model was applied in \cite{puk1} to intraday returns for the USD/JPY and USD/EUR exchange rates during stress periods. Instead of our quartic potential with stable coefficients, only a cubic potential was measured. Morevoer, its coefficients, including their signs, were found to rapidly vary in time. \\

However, these studies were based on very different data sets, namely tick data (instead of daily data) for single assets over time periods of several weeks (instead of decades). Thus, it is no surprise that the stable quartic potential (corresponding to the cubic trem in (\ref{cubic})) was not found in \cite{puk1}: as we have seen, in order to detect it with strong statistical significance, one needs not only decades of data, but also aggregate them over a broadly diversified set of assets. Also, since the coefficient of the cubic potential reported in \cite{puk1} varies rapidly in time, it can be expected to average out over long time scales. This is consistent with the fact that we do not observe a cubic potential in our empirical long-term analysis.

\section{Summary and Discussion} 

In this paper, we have empirically observed the interplay of trends and reversion in financial markets, based on 30 years of daily futures returns across equity indices, interest rates, currencies and commodities. We have considered trends over ten different time horizons of $T=2^k$ days with $k\in \{1,2,...,10\}$, ranging from 2 days to approximately 4 years. For a given market $i$ on a given day $t$, we have defined the trend strength $\phi_{i,k}(t)$ as the statistical significance ($t$-statistics) of a smoothed version of its mean return over the past $2^k$ days, in excess of the market's long-term risk premium.\\

Our key results, as illustrated in figs 2 and 3, are the following: for a given market $i$ and each time horizon labeled by $k$, tomorrow's normalized log-return $R_{i}(t+1)$ can accurately be modeled by a cubic polynomial of today's trend strength in that market:
\begin{equation}
R_i(t+1)=\alpha_i+b\cdot f_k\cdot  \phi_{i,k}(t)+c \cdot \phi_{i,k}^3(t)+\epsilon_i(t+1).\label{model}
\end{equation}
Here, $\epsilon_i$ represents random noise. $\alpha_i$ is the normalized long-term risk premium of market $i$, which has not been not the focus of this paper. Instead, we have concentrated on determining the coefficients $b, c,$ and the function $f_k$, which measure how the expected return of an asset varies in time. As discussed, we interpret $b$ as the persistence of trends, and $c$ as the strength of trend reversion. Within the limits of statistical significance, we find that they are universal, i.e., the same for all assets. Over the past 30 years, we find from table 4:
\begin{equation}
b\sim +2.0\%\ ,\ \ \ c\sim -0.6\%\ \label{coef}
\end{equation}
While the strength of reversion is approximately constant, we find that the persistence of trends depends on the time horizon of the trend. Within the range of time scales considered here, it can be approximated by a parabolic function of the log of the time scale:
\begin{equation}
f_k\sim 1-{(k-k_0)^2\over\Delta k^2}\ \ \ \text{with}\ \ \ k_0\sim 6\ ,\ \ \ \Delta k\sim 5.\label{para}
\end{equation}
This implies that trends may only be stable if the log of the time horizon is within the range $k_0\pm\Delta k$, corresponding to time scales from a few days to several years. The parameters $k_0$ and $\Delta k$ are also universal. By bootstrapping and cross-validation, we have found that all four parameters in (\ref{coef}) and (\ref{para}) are statistically highly significant out-of-sample. \\

Let us now discuss these results. First, they imply that trends tend to revert above a critical trend strength, where the linear and cubic term in (\ref{model}) balance each other. This critical trend strength lies below 2 in all cases. In other words, by the time a trend has become statistically significant, such that it is obvious in a price chart, it is already over. This supports a variant of the efficient market hypothesis \cite{mandel,samu,fama}: inefficiencies in financial markets are eliminated before they become strongly statistically significant. \\

Despite being insignificant, small trends can add value for investors through tactical asset allocation strategies, if accompanied by appropriate risk management and broad diversification across assets. While this paper does not recommend investment strategies, we note that the inclusion of the cubic term in (\ref{model}) appears to be a major improvement over classical trend-following, as it takes investors out of trends before they are likely to revert (see also the comments on systematic asset management in appendix A3). We believe that publishing such strategies and subjecting them to an academic discussion and independent review will ensure a high level of professionality in asset management. \\

Trend-following has been very successful in the 80's and 90's, when it was the proprietary strategy of a limited number of traders. By now, large amounts of capital have flown into this strategy, so it can no longer be expected to provide a “free lunch.” Indeed, while we have not observed a consistent weakening of the strength of reversion $c$, we have seen that the persistence of market trends $b$ has clearly decreased over the decades. This measures the rate, at which markets are becoming more efficient with respect to trends.\\

What will happen, when all investors try to exploit trends and reversion? Then both phenomena should weaken, until they earn a moderate equilibrium return that just compensates for the systematic risk of these strategies and their implementation costs. In this sense, trend-following and mean reversion may just become "alternative market factors" as part of the general market portfolio. In fact, the weakening of $b$ that we have observed here indicates that this development is already well underway at least for traditional trend-following.\\

On a conceptual level, our precise measurement of trends and reversion reveals intriguing analogies with critical phenomena in physics. They support the conjecture that financial markets can be modeled by statistical mechanical systems near second-order phase transitions. In such a model, Buy/Sell orders would represent microscopic degrees of freedom that live on a "social network" of traders. The trend strength would play the role of an order parameter, whose dynamics is described by the stochastic differential equations (\ref{stoch},\ref{lange}). Together with an extension of the scaling behavior (\ref{para}) to shorter and longer horizons, these equations provide an empirical starting point for developing such a model.\\

If such a statistical mechanical theory of financial markets can be established, it will introduce powerful concepts from field theory into finance, such as the renormalization group, critical exponents, and Feynman diagrams. This will lead to a new and deeper understanding of financial markets, and phenomena such as trends, reversion, and shocks will become more accessible to scientific analysis. Further research in this direction is underway.

\section*{Acknowledgements} 

I would like to thank W. Breymann for discussions and encouragement to publish these observations, and A. Ruckstuhl for advice on statistical matters. I would also like to thank J. Behrens for  many interesting conversations and cooperation at Syndex Capital Management from 2008-2014, where some of these observations were initially made. This research is supported by grant no. CRSK-2 190659 from the Swiss National Science Foundation.

\newpage
\section*{Appendix} 

The following table compares the regression results of subsection 4.2 with the results that would be obtained for alternative choices of some of the parameters:\\

\begin{center}
\begin{tabular}{ |p{2cm}||p{1.3cm}|p{1.3cm}|p{1.3cm}|p{1.3cm}|p{1.3cm}||p{1.3cm}|p{1.3cm}|p{1.3cm}| }	\hline
 	{\it Table 6} & \multicolumn{5}{|l||}{Different Caps/Floors}& \multicolumn{3}{|l|}{Inclusion of Risk Premia}\\ \hline
	{Cap/Floor}& 2.0 &2.25  &{\bf 2.5}&2.75 &3.0&{\bf 0\%} &50\%&100\%\\	\hline
$b$	&	2.05\%	&	2.03\%	&	2.00\%	&	1.95\%	&	1.88\%	&	2.00\%	&	2.18\%	&	2.32\%	\\	
$t$-stat.	&	4.0	&	4.0	&	4.2	&	3.7	&	3.6	&	4.2	&	4.4	&	3.3	\\	\hline
$c$	&	-0.75\%	&	-0.69\%	&	-0.63\%	&	-0.58\%	&	-0.53\%	&	-0.63\%	&	-0.62\%	&	-0.53\%	\\	
$t$-stat.	&	2.7	&	2.8	&	2.6	&	2.5	&	2.4	&	2.6	&	2.6	&	2.9	\\	\hline
$k_0$	&	5.90	&	5.85	&	5.78	&	5.72	&	5.69	&	5.78	&	6.75	&	8.21	\\	
$\Delta k$	&	4.87	&	4.90	&	4.87	&	4.77	&	4.67	&	4.87	&	5.96	&	7.20	\\	\hline
$R^2_{adj}$	&	1.12	&	1.39	&	1.64	&	1.82	&	1.96	&	1.64	&	1.64	&	1.59	\\	
$R^2_{adj}$ aggr.	&	0.74	&	0.98	&	1.22	&	1.40	&	1.54	&	1.22	&	1.19	&	1.22	\\	\hline
\end{tabular}
\end{center}\vspace{5pt}

\subsection*{A1. Caps and Floors for the Trend Strength} 

In section 2, we have capped the magnitude of the trend strength at 2.5 to limit the effect of outliers on the results. Table 6 (col. 2-6) compares the results of section 4.2 for the alternative caps/floors of 2.0, 2.25, 2.5, 2.75, and 3.0. We observe the following:
\begin{itemize} 
\item
Increasing the cap beyond 2.5 increases the adjusted R-squared, but decreases the significance (t-statistics) of the regression betas. This makes sense intuitively, because the results are now dominated by the regime of strong reversion at "outlier" trend strength $\vert\phi\vert>2.5$. As such outliers are rare, the statistical significance decreases.
\item
Decreasing the cap below 2.5 decreases the adjusted R-squared without improving the overall significnce of the regression betas. This is also understandable, as it removes much of the reversion regime from the analysis. Thus, our cap/floor of $\pm$2.5 is a good compromise, where neither trends nor reversion outliers dominate the results. 
\end{itemize}

\subsection*{A2. Long-term Risk Premia} 

In equation  (\ref{phi}), we have removed the long-term risk premia $\mu_i$ from the trend strengths $\phi_{i,T}$. Here, we explain what happens if we do {\it not} remove the risk premia from the trend strengths: 
\begin{itemize} 
\item Trends in markets, for which such risk premia are generally assumed, would then have an upward bias, i.e., positive expectation value. Especially very-long-term trends in equity and bond markets would almost always be positive and never revert.
\item Table 6 (col. 7-9) shows how this mix-up of trends and risk premia would modify the results of subsection 4.2 (shown under "0\%"). If 50\% or 100\% of the risk premia were included in the definition of the trend strength, the parameter $k_0$ (which measures the time horizon at which the persistence $b$ of trends peaks) would strongly increase.
\item In fact, if we think of risk premia as  trends with infinite time horizon, we expect that, without removing risk premia, the trending regime of sub-section 4.2 would extend all the way to infinite horizon. We would then model it by a parabola instead of an ellipse. 
\end{itemize}

\subsection*{A3. Comments on Systematic Asset Management} 

The key motivation for this article is to lay the empirical basis for a statistical-mechanical model of financial markets, which can hopefully explain the analogies with critical phenomena in physics. Nevertheless, let us briefly comment on implications for systematic trading:
\begin{itemize}\addtolength{\itemsep}{-1.5 pt}
\item According to the back-of-the-envelope estimate of sub-section 3.3, an annualized Sharpe ratio of order 1 for a systematic futures trading strategy corresponds to an adjusted R-squared of about 4 basis points in predicting daily returns of individual markets.
\item By the same argument, the aggregated adjusted R-squared of 6 basis points of sub-section 4.2. corresponds to an annual Sharpe ratio of $\sqrt{1.5}$ for a {\it market-neutral} strategy. 
\item Trading costs reduce the Sharpe ratio, especially at intra-month time horizons. Diversi-fying into new types of markets or including risk premia can increase the Sharpe ratio. 
\item Risk control mechanisms, such as sizing positions based on current market volatility, or stop-losses in the reversion regime, can either {\it decrease} or {\it increase} the Sharpe ratio.
\end{itemize}

\end{document}